\documentstyle[preprint,tighten,floats,prd,aps]{revtex}
\input epsf

\newcommand{\be}{\begin{eqnarray}}
\newcommand{\ee}{\end{eqnarray}}

\begin{document}
\draft
\preprint{\vbox {\hbox{UAHEP9812}
\hbox{November 1998}
\hbox {hep-ph/9812340} } } 
\title{Light Gluino Predictions for Jet Cross Sections \\in Tevatron
Run II}
\author{L. Clavelli\footnote{lclavell@bama.ua.edu}} 
\address{Department of Physics and Astronomy, University of Alabama,\\
Tuscaloosa AL 35487}
\maketitle
\begin{abstract}
The CDF inclusive jet transverse energy cross section at $1.8 TeV$
suggests anomalous behavior at both low and high transverse energies.
In addition the scaled ratio of the $0.63 TeV$ to $1.8 TeV$ data lies
significantly below the standard model prediction and suggests
structure not attributable to standard model processes. These
anomalies are in line with what would be expected in the light gluino
scenario. We perform a unified fit and extrapolate to two TeV to
predict the results at run II.
\end{abstract}

The CDF collaboration at Fermilab has published \cite{Abe1} a study of
the jet inclusive transverse energy cross section in $p {\overline p}$
cross sections at $1.8 TeV$ which suggest the possibility of anomalous
behavior in both the low and high transverse energy regions. D0 has
not published results in the low transverse energy region but has
presented data at high transverse energy which appear to be consistent
with either the CDF result or the standard model. The apparent anomaly
at high $E_T$ seen by CDF could, therefore, be a statistical
fluctuation. It has also been suggested
\cite{Lai} that these results are compatible with the standard model if
the gluon distribution at high x is appreciably higher than expected on
the basis of previous fits. On the other hand, the anomalous behavior
observed by CDF in both the low and high $E_T$ regions is also
consistent with that expected if the gluino of supersymmetry is light
(below $10 GeV$ in mass is sufficient) \cite{CT}. Although all direct
searches for a light gluino have turned up negative, many indirect
indications of such a light color octet parton have been noted. A
partial list is contained in the references of \cite{CT}.

The measured inclusive cross section at center of mass energy $\sqrt
s$ to produce a jet of transverse energy $E_T$ averaged over a certain
rapidity interval is theoretically expected to have the form
\be
 d\sigma/dE_T  = \alpha_s(\mu)^2 s^{-3/2} F(X_T,\frac{\Lambda}{\sqrt s},
     \frac{m}{\sqrt s})
    + {\cal{O}}(\alpha_s^3)
\ee
Here $\mu$ is the scale parameter, $X_T=2E_T/\sqrt s,\quad \Lambda$ is
the QCD dimensional transmutation parameter, and m represents any of
the masses of the strongly interacting particles in the theory. Taken
to all orders the cross section is independent of $\mu$ but at finite
order the theoretical result depends on $\mu$ which must therefore be
treated as a parameter of the theory. The CDF best fits correspond to
$\mu=E_T/2$. At high energy the scaling function F depends only on
$X_T$ . The CDF data for this cross section compared to the
next-to-leading order (NLO) QCD predictions are below unity at low
$E_T$ and rise dramatically above unity at high $E_T$. In the
Supersymmetry (SUSY) treatment of \cite{CT2} this behavior was
attributed to three phenomena.
\begin{itemize}
\item  With a light gluino the strong coupling constant runs more slowly
being higher than the standard model at high $\mu$ and lower at low
$\mu$.
\item The production of gluino pairs increases F by a roughly
uniform factor of 1.06 for all $E_T$
\item A squark, if present, will cause a bump in the cross section at
about ${m_{\tilde Q}}/2$.
\end{itemize}

The fit of \cite{CT2} used the CDF suggested value of $\mu$ and a
value of $\Lambda$ corresponding to $\alpha_s(M_Z)=0.113$ and a squark
mass of about $106 GeV$. The theoretical ratio of the SUSY prediction
relative to the standard model prediction is relatively insensitive to
higher order corrections since both will have roughly equal higher
order enhancements. In this work the CTEQ3 parton distribution
functions (pdf's) were used. In a later study \cite{Bhatti}, CDF
considered the scaled ratio of the inclusive jet $E_T$ cross sections
at $630 GeV$ and $1.8 TeV$.
\be
  r(X_T) = \frac{s^{3/2} d\sigma/dE_T \quad(\sqrt{s} = 630 GeV)}{
              s^{3/2} d\sigma/dE_T \quad(\sqrt{s} = 1800 GeV)}
\ee

Since at both energies, $\sqrt{s}$ is much greater than the QCD scale
parameter $\Lambda$ and all the quark masses of the standard model
(except the top quark which contributes negligibly at these energies),
the standard model prediction modulo residual corrections from higher
order and from scaling violation in the pdf's is just
\be
        r(X_T) = \frac{\alpha_s^2(\lambda X_T \cdot 0.630GeV/2)}
              {\alpha_s^2(\lambda X_T \cdot 1.8 TeV/2)}
\ee

We have assumed here that the appropriate choice of $\mu$ is $\lambda
E_T$ with $\lambda=1/2$ being the result of the CDF best fit to the
$1.8 TeV$ data. The full standard model prediction with corrections
incorporated seriously overestimates the CDF data. In addition there
is a possible structure in r that, if real, might suggest the
existence of a strongly interacting particle in the $100 GeV$ region
with a production cross section many times larger than that of top. As
always, there is the possibility that the anomaly is due to systematic
errors although it would be surprising if such errors induced
structure in $E_T$. In fact, the D0 experiment does not confirm the
existence \cite{D0} of structure in r suggesting, therefore, an
explanation in terms of systematic errors. Although the systematic
errors could easily affect the normalization of the r parameter, it
would be surprising if they affect the point to point errors. These
systematic errors derive primarily from the lower energy (630 GeV)
data and hence the existence or non-existence of structure should be
definitively resolved by comparing the ratio of the $2 TeV$ data which
will be available beginning in the year 2000 with the $1.8 TeV$ data.
Although the energy step is small, the greatly increased luminosity in
run II coupled with the small systematic errors in the $1.8 TeV$ data
should guarantee sufficient sensitivity to settle the question.

The features observed by CDF in the scaling ratio are those expected
in the light gluino scenario \cite{CT2}. The slower fall-off of
$\alpha_s$ predicts that the r parameter should be generally lower
than the standard model expectations in agreement with the data. In
addition a squark in the $100 GeV$ mass range would provide a bump in
each cross section at roughly fixed $E_T={m_{\tilde Q}}/2$. This would
lead to a dip-bump structure separated by a factor of 1.8/0.63 in
$X_T$ in qualitative agreement with the CDF data. If the bump had
occured at lower $X_T$ than the dip there would have been no
possibility of a fit in any model where the structure was attributed
to a new particle. Reference \cite{CT2} provided two fits to the CDF
data. The first used the CTEQ3 pdf's and the scale choice $\mu=E_T/2$
with a squark mass of $130 GeV$. In the CTEQ3 pdf's there are, of
course, no initial state gluinos so the cross section bump derives
from the reaction
\be
                   q g \rightarrow q {\tilde g} {\tilde g}
\label{qg}
\ee
with an intermediate squark in the $q {\tilde g}$ channel. The
dynamics are such that the initial state gluon splits into two
dominantly collinear gluinos one of which interacts with the initial
state quark to produce the intermediate squark. Other non-resonant
light gluino contributions to the cross section come from the parton
level processes
\be
      q {\overline q} \rightarrow {\tilde g} {\tilde g}\\
      g g \rightarrow {\tilde g} {\tilde g}
\ee

If the gluino is light it should have a pronounced presence in the
proton dynamically generated from the gluon splitting discussed above.
Two groups \cite{BB,RV} have analyzed deep inelastic scattering
allowing for a light gluino and presented fits to the gluino pdf as
well as modifications of the other pdf's due to the gluino presence.
In ref. \cite{CT2} we compared the scaling violation using the
R\"uckl-Vogt pdf's with that using the CTEQ3 set. With intrinsic
gluinos there are extra contributions to the jet inclusive cross
sections from the processes
\be
       g {\tilde g} \rightarrow g {\tilde g} \\
       q {\tilde g} \rightarrow q {\tilde g} \\
       {\tilde g} {\tilde g} \rightarrow {\tilde g}{\tilde g}\\
       {\tilde g} {\tilde g} \rightarrow g g\\
       {\tilde g} {\tilde g} \rightarrow q {\overline q}
\ee
The second process replaces the higher order reaction of Eq.\
(\ref{qg}) and provides a direct channel pole at the mass of the
squark leading to a peak in the transverse energy cross sections. We
treat the squark as a resonance in the quark-gluino channel. Each of
these reactions of course is subject to higher order corrections but
these tend to cancel in the scaling ratio and in the ratio of the SUSY
transverse energy cross section to that of the standard model. In this
second fit the scale $\mu$ was chosen to be the parton-parton CM
energy.

The purpose of the current work is to return to the inclusive
jet transverse energy cross section and seek a combined fit to this
plus the scaling curve allowing for intrinsic gluinos in the proton.
Fitting both the scaling curve $r(X_T)$ and the $1.8 TeV$ cross
section is equivalent to fitting the transverse energy cross section
at both $1.8 TeV$ and $0.63 TeV$. Using the parameters of this
combined best fit we then present the predictions for the $E_T$ cross
section at $2 TeV$ CM energy of run II and the scaling curve for $1.8
TeV / 2 TeV$. The primary parameters of the combined fit are the scale
$\mu$, the QCD $\Lambda$ parameter or equivalently $\alpha_s(M_Z)$,
and the squark mass $m_{\tilde Q}$. We find the optimal values
\be
     \mu = 0.6 E_T\\
     \alpha_s(M_Z) = 0.116\\
     m_{\tilde Q} = 133 GeV
\ee
In the fit we estimate NLO corrections by the K factor $1 +10
\alpha_s(\mu) /\pi$ and we simulate resolution smearing by increasing the
width of the squark by a factor of 2 from its SUSY QCD prediction $2
\alpha_s m_{\tilde Q}/3$. In addition it is known that the systematic
errors in the $630 GeV$ data form a fairly broad band \cite{Geer}. We
therefore allow the scaling data to float by a uniform factor near
unity. The results are presented in figures 1-4.

Figure 1 shows the
fit to the $1.8 TeV$ jet inclusive $E_T$ cross section averaged over
the CDF rapidity range $0.1 < |\eta| < 0.7$. In order to compare with
the data of ref. \cite{Bhatti}, the light gluino prediction is
plotted relative to the QCD prediction given to us in a private
communication by the author of that reference. At high $E_T$ the fit
goes through the lower range of the CDF errors which suggests it is
also consistent with the D0 data. The fit qualitatively reproduces the
dip at low $E_T$ and shows a peak at low $E_T$ due to the $133 GeV$
squark. Figure 2 shows the scaling function $r(X_T)$ as given in the
light gluino scenario with a $133 GeV$ squark and as given by the
standard model. The data has been moved up by a uniform factor of 1.2
which is consistent with the effect of systematic errors in the $630
GeV$ data. The height and width of the dip-bump structure is in
qualitative agreement with the expectations of the light gluino plus
$133 GeV$ squark model. One might expect that a full simulation
including hadronization and detector acceptance could somewhat shift
this mass.

If a squark exists at $133 GeV$ it should be apparent in
the $e^+e^-$ annihilation cross section through the
quark-squark-gluino final state \cite{CCFHPY}.
The L3 data \cite{L3} shows what is possibly an upward
statistical fluctuation in the hadronic cross section in $e^+ e^-$
annihilation in the $130 GeV$ region. Since the gluino decays are
expected to leave very little missing energy, the quark-squark-gluino
final state might also explain an apparent surplus in the visible
energy cross section at high $E_{vis}$ \cite{L32}. In addition, a SUSY
symmetry breaking scale of $133 GeV$ would, in the light gluino
scenario, predict stop quarks in the region just above the top and
could explain some anomalies in the top quark events
\cite{CG} and lead to an enhancement in the deep inelastic cross section
at high $Q^2$ and high hadronic mass \cite{CC}. If there is indeed a
light gluino and a squark in the $100 \sim 135 GeV$ region, a dip-bump
structure should also be found at Lep II in the scaling ratio of the
inclusive dijet cross section in $e^+ e^-$ annihilation.
\be
        r(M^2/s) = \frac{ s^{2} d\sigma/dM^2 \quad (\sqrt{s}=E_1)}
                    { s^{2} d\sigma/dM^2 \quad (\sqrt{s}=E_2)}
\ee
where both $E_1$ and $E_2$ are above the squark mass. Since the squark
decays in the present model into quark plus gluino, the excess should
be in the four-jet sample but should not appear in the pair production
of two high mass states.

In figure 3, we show the predictions for the
jet inclusive $E_T$ cross section in $p {\overline p}$ collisions at
the energy $2 TeV$ relative to the standard model expectations. The
curve shows a pronounced peak at ${m_{\tilde Q}}/2$ and is generally
$5$ to $10 \%$ below unity due to the slower running of $\alpha_s$ in
the light gluino case and to the scaling violations in the parton
distribution functions. In Figure 4 the scaling ratio
\be
  r(X_T) = \frac{s^{3/2} d\sigma/dE_T \quad(\sqrt{s} = 1.8 TeV)}{
              s^{3/2} d\sigma/dE_T \quad(\sqrt{s} = 2 TeV)}
\ee
is plotted for the case of light gluino plus $133 GeV$ squark and for
the case of light gluino but no squark present (non-resonant solid
line). The dash-dotted curve gives the prediction of the standard
model.

Although we have not attempted to estimate hadronization
corrections nor resolution smearing (apart from doubling the squark
width), we expect that run II will be sensitive to the predicted peaks
if they exist and will therefore either discover or rule out a squark
in the $100 GeV$ mass region in conjunction with a light gluino. With
additional information on dijet mass and angular distributions
\cite{TC,T,DHR}, the Run II measurements are sensitive to a light
gluino with a squark up to $1 TeV$. Since most of the value of SUSY
would be lost with squarks so high in mass, Run II should definitively
settle the question as to whether the light gluino indications
including those referenced in \cite{CT} are the first signs of SUSY or
merely an amazing string of coincidences attributable to systematic
errors.

\acknowledgements
This work was supported in part by the US Department of Energy under
grant no. DE-FG02-96ER-40967. We acknowledge useful conversations on
the Fermilab data with I. Terekhov.

\break
\begin{figure}
\hskip 4.0cm
\epsfxsize=3in \epsfysize=3in \epsfbox{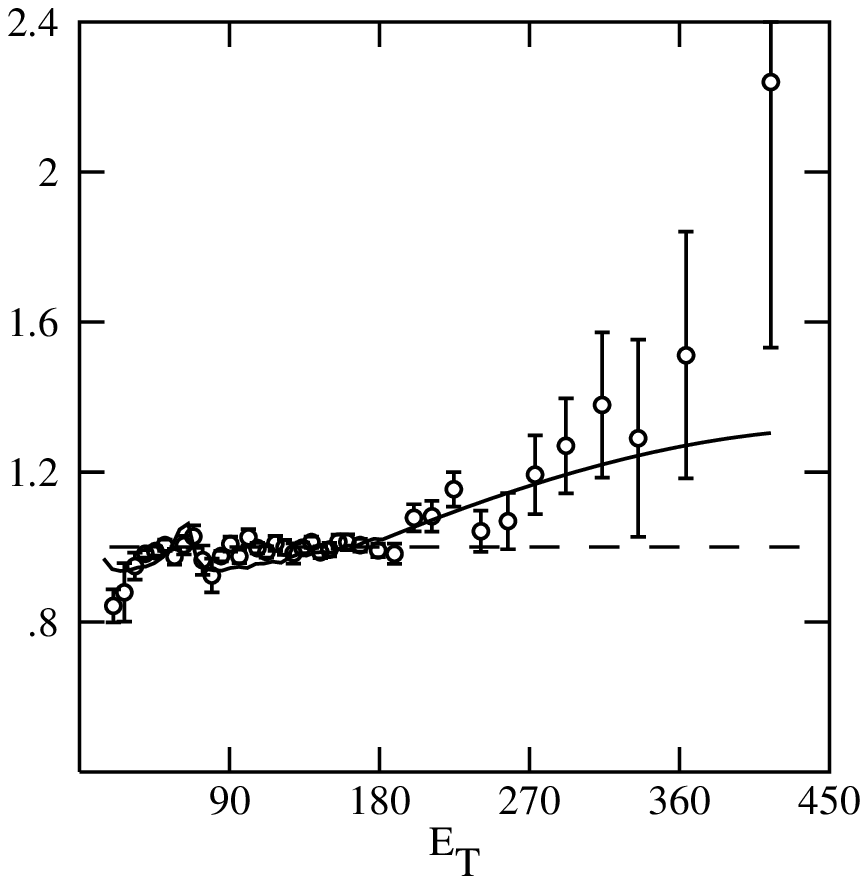}
\caption{Ratio of the inclusive jet transverse energy cross section
with a light gluino and a $133 GeV$ squark to that of the standard
model. CDF data at $1.8 TeV$ is superimposed.}
\label{Fig. 1}
\end{figure}
\begin{figure}
\hskip 4.0cm
\epsfxsize=3in \epsfysize=3in \epsfbox{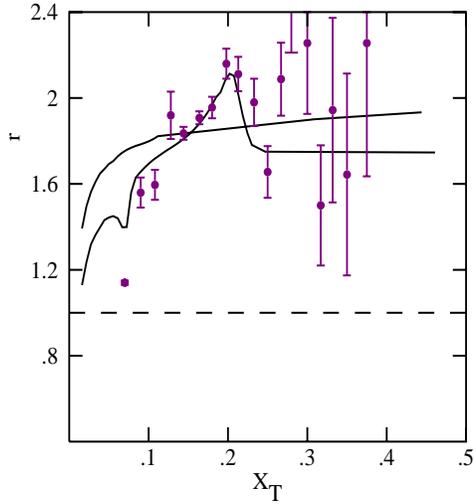}
\caption{CDF data for the scaling ratio of the inclusive jet transverse
energy cross section at $630 GeV$ relative to $1.8 TeV$ compared to
the fit with a light gluino plus $133 GeV$ squark and to the standard
model prediction (structureless curve). The data has been moved up by
$20 \%$ consistent with the systematic errors in the $630 GeV$ data.}
\label{Fig. 2}
\end{figure}
\begin{figure}
\hskip 4.0cm
\epsfxsize=3in \epsfysize=3in \epsfbox{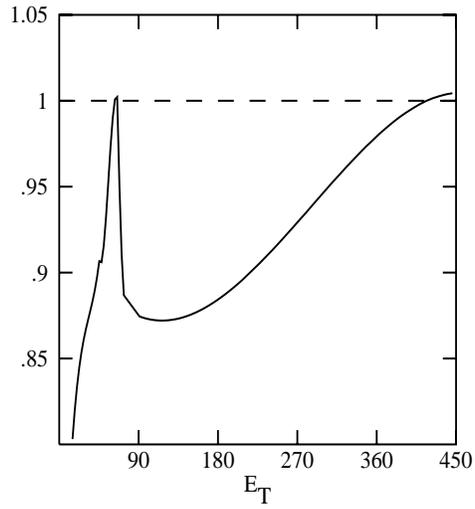}
\caption{Predicted ratio of the inclusive jet transverse energy cross
section with a light gluino and a $133 GeV$ squark to that of the
standard model for $2 TeV p {\overline p}$ collisions.}
\label{Fig. 3}
\end{figure}
\begin{figure}
\hskip 4.0cm
\epsfxsize=3in \epsfysize=3in \epsfbox{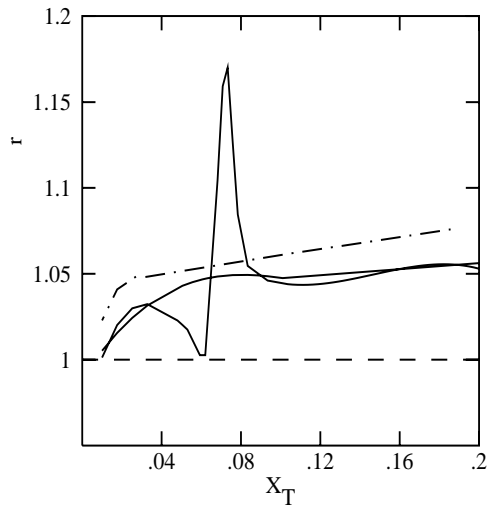}
\caption{predicted scaling ratio of the inclusive jet transverse
energy cross section at $1.8 TeV$ relative to $2 TeV$ for a) light
gluino only (no squark) and b) light gluino plus $133 GeV$ squark. The
dash-dotted curve shows the standard model prediction.}
\label{Fig. 4}
\end{figure}
\end{document}